\documentclass[sigconf,10pt]{acmart}

\usepackage{url}
\usepackage{xspace}
\usepackage{etoolbox}
\usepackage{comment}
\usepackage{booktabs}
\usepackage{graphicx}
\usepackage{makecell}
\usepackage{multirow}
\usepackage{subcaption}
\usepackage{tikz}
\usepackage{caption}
\usepackage{flushend}
\usepackage[shortcuts]{extdash}
\usepackage{balance}
\usepackage{paralist}
\usepackage{enumitem}
\usepackage{siunitx}
\usepackage{listings}
\usepackage{xcolor}
\usepackage[capitalize,nameinlink]{cleveref}
\usepackage[most]{tcolorbox}
\usepackage{listings}
\usepackage{soul}
\usepackage{pgfplots}
\pgfplotsset{compat=1.18}
\mathchardef\mhyphen="2D 

\newcommand{\eg}{\emph{e.g.,}\xspace}

\newcommand{\myparagraphnodot}[1]{\vspace{\smallskipamount}\noindent\textbf{#1\xspace}}

\usepackage[most]{tcolorbox}
\tcbset{textmarker/.style={%
        enhanced,
        parbox=false,boxrule=0mm,boxsep=0mm,arc=1.5mm,
        outer arc=1.5mm,left=2mm,right=2mm,top=4pt,bottom=3pt,
        toptitle=1mm,bottomtitle=1mm,oversize}}

\newtcolorbox{simplenoteBox}{colback=white, colframe=black, boxrule=0.2mm, arc=0mm, auto outer arc, boxsep=0mm, left=2mm, right=2mm, top=1mm, bottom=1mm} 
\newtcolorbox{noteBox}{textmarker,
    colback=gray!8!white}
\newcommand{\takeaway}[1]{\begin{noteBox} \textbf{Takeaway:} #1 \end{noteBox}}

\newtoggle{showmarks}
\toggletrue{showmarks}

\iftoggle{showmarks}{
  
  \newcommand{\oldtext}[1]{\leavevmode\textcolor{red}{OLD: #1}}
  
  \newcommand\gic[1]{\textcolor{olive}{GIC: #1}}
}{
  
  \newcommand{\oldtext}[1]{\unskip}
  
  \newcommand\gic[1]{\unskip}
}

\definecolor{redbg}{RGB}{255,220,220}
\definecolor{greenbg}{RGB}{220,255,220}
\definecolor{redframe}{RGB}{200,100,100}
\definecolor{greenframe}{RGB}{100,180,100}
\definecolor{graytext}{RGB}{100,100,100}

\lstdefinestyle{promptstyle}{
    basicstyle=\ttfamily\small,
    breaklines=true,
    frame=none,
    aboveskip=0pt,
    belowskip=0pt,
    escapeinside={(*@}{@*)},
}
\tcbuselibrary{listings,skins,breakable}

\begin{document}

\title{Benchmarking Compound AI Applications for Hardware-Software Co-Design}

\author{Paramuth Samuthrsindh}
\authornote{Equal contribution}
\affiliation{\institution{MIT CSAIL}}
\author{Angel Cervantes}
\authornotemark[1]
\affiliation{\institution{MIT CSAIL}}
\author{Varun Gohil}
\authornote{Corresponding authors: \{varuncg, girfan\}@mit.edu}
\affiliation{\institution{MIT CSAIL}}
\author{Gohar Irfan Chaudhry}
\authornotemark[2]
\affiliation{\institution{MIT CSAIL}}
\author{Christina Delimitrou}
\affiliation{\institution{MIT CSAIL}}
\author{Adam Belay}
\affiliation{\institution{MIT CSAIL}}

\begin{abstract}
Compound AI applications, composed from interactions between Large Language Models (LLMs), Machine Learning (ML) models, external tools and data sources are quickly becoming an integral workload in datacenters. 
Their diverse sub-components and use-cases present a large configuration-space across the deployment stack---ranging from applications and serving software down to hardware---each of which may influence the application performance, deployment cost, and/or resource consumption.
Despite their rapid adoption, however, the systems community lacks a standardized benchmark for analyzing this complicated design-space and guiding in system design.
In this work, we present our benchmarking suite used for \emph{cross-stack analysis of Compound AI applications}.
Using this, we derive key takeaways and design principles spanning several layers of the stack for hardware-software co-design to unlock higher resource-efficiency.
\end{abstract}

\renewcommand\footnotetextcopyrightpermission[1]{}
\settopmatter{printacmref=false}
\maketitle
\pagestyle{plain}

\section{Introduction}

The rapid ascent of Large Language Models (LLMs) has fundamentally reshaped the landscape of artificial intelligence (AI), transitioning from niche research tools to the backbone of modern software.
As these models gain robustness and reliability across diverse domains ranging from healthcare~\cite{goyal2024healai} and education~\cite{dai2024agent4edu} to software engineering~\cite{cursorFeatures2025,zhang2024pybench}, they are no longer being deployed as isolated entities.
Instead, we are witnessing the rise of Compound AI applications---complex architectures that combine LLMs with multiple sub-components to solve complex tasks~\cite{compound-ai-blog}.

These Compound AI applications function similarly to microservice-style applications, orchestrating interactions between various entities to reach a final output.
Typically, an LLM acts as the central reasoning engine, but the system is comprised of many other sub-components, including specialized ML models like speech-to-text, external tools, such as web-search engines or code interpreters, and data sources like vector databases.
While this modularity enables immense flexibility, it also introduces a massive configuration space where application semantics, software stack settings, accelerator choices, and hardware-level power management all collide to dictate end-to-end performance~\cite{chaudhry2025murakkabresourceefficientagenticworkflow,chaudhry2025towardsresourceefficientcompoundaisystems}.

Historically, the systems community has relied on standardized benchmarks to navigate similar shifts in computing, using SPEC~\cite{henning2006speccpu} for general-purpose processing and DeathStarBench~\cite{gan2019deathstarbench} for the transition to microservices.
The machine learning community has also established MLCommons~\cite{mattson2020mlperftrainingbenchmark,chung2025mlenergybenchmarkautomatedinference} to provide vital hardware-software analysis.
However, existing benchmarks almost exclusively focus on single-model inference and training while overlooking the trade-offs unique to applications that consist of interactions between multiple models and tools executing on heterogeneous hardware.
This leaves a gap in the benchmarking landscape that focuses on Compound AI applications and their trade-offs.

\myparagraphnodot{Our Work:}
We present a \emph{cross-stack benchmark suite for Compound AI applications}.
Using this, we analyze the trade-offs stemming from configurations across \emph{all levels} of the serving stack including application, software, and the hardware platform level.
In this work, we analyze three representative applications from the benchmark suite focusing primarily on the following aspects:
(a) hardware selection and its implications on cost, performance, and energy,
(b) configuration tuning and its implications on power consumption patterns, and
(c) application semantics and their implications on cache usage and routing policies.
Based on our analysis, we present important takeaways for resource-efficient system design and hope to provide a better understanding of performance characteristics to application developers and reveal opportunities for better hardware-software co-design to cloud providers.

We make the following contributions:
\begin{enumerate}[leftmargin=*]
    \item We introduce and study two representative compound AI applications---Video-QA~\cite{zhang2024omagent}, OpenEvolve~\cite{openevolve}, and Retrieval-Augmented Generation (RAG)~\cite{lewis2020rag}---to analyze system-level characteristics across heterogeneous CPU and GPU resources.
    \item We identify that Compound AI systems lack a ``one-size-fits-all'' hardware setup, as different sub-components exhibit varying sensitivities to frequency scaling; consequently, we demonstrate that system efficiency can be maximized only by dynamically adjusting GPU configurations and power management strategies based on the specific load and component-level bottlenecks of the workflow.
    \item We demonstrate that dynamic prompt restructuring and cache-aware routing can significantly boost cache hit rates, reducing end-to-end latency by up to 23.8\% and energy usage by up to 12\%.
\end{enumerate}

The benchmark will be made open-source.
\footnote{\url{https://github.com/mit-caisys/caisys-bench}}
\section{Benchmark Design}

\begin{figure*}[t]
    \centering
    \begin{subfigure}[t]{0.33\textwidth}
        \centering
        \includegraphics[width=\linewidth]{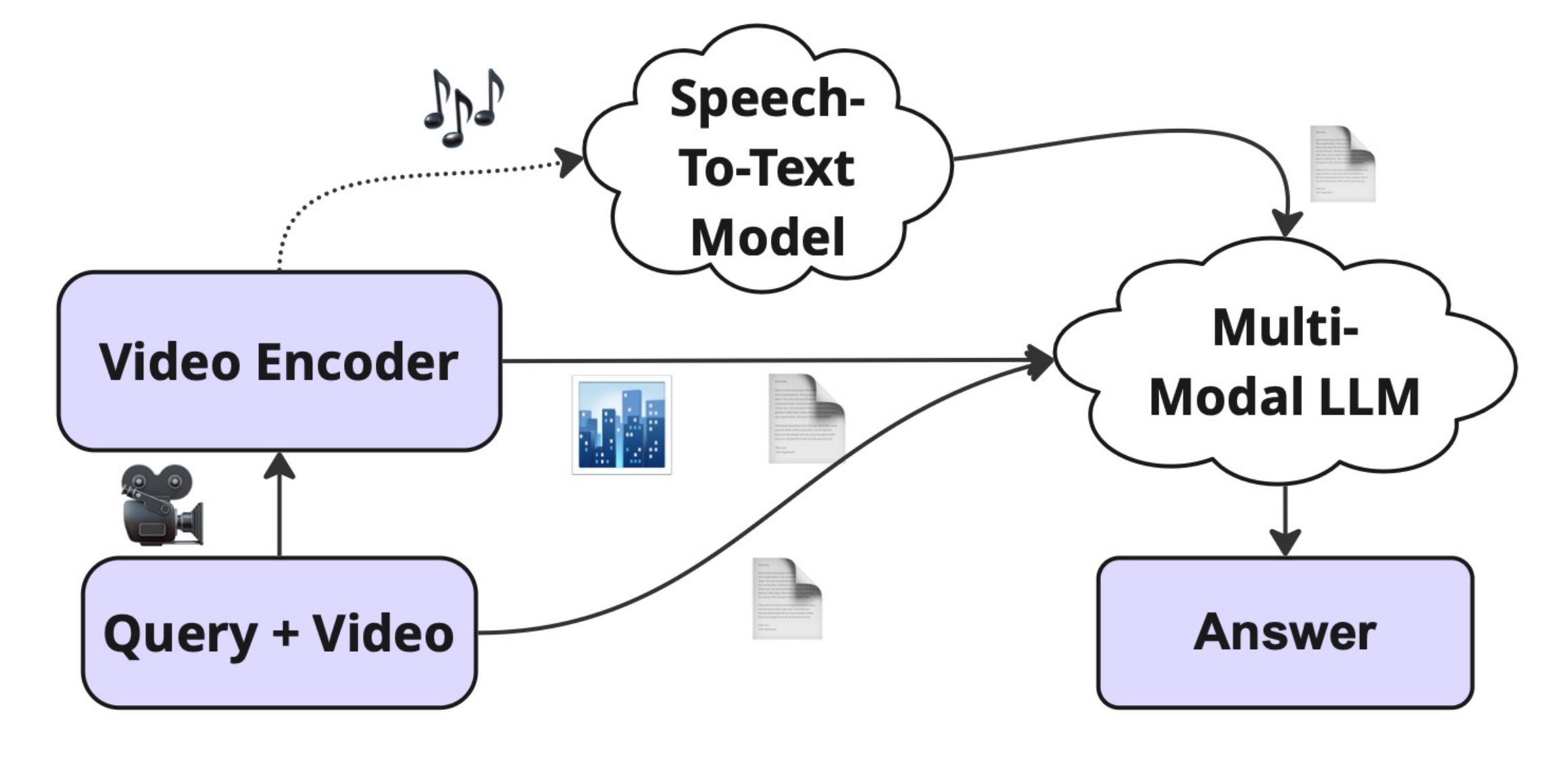}
        \caption{Video Question/Answering.}
        \label{fig:video-qa_workflow}
    \end{subfigure}%
    \hfill
    \begin{subfigure}[t]{0.33\textwidth}
        \centering
        \includegraphics[width=\linewidth]{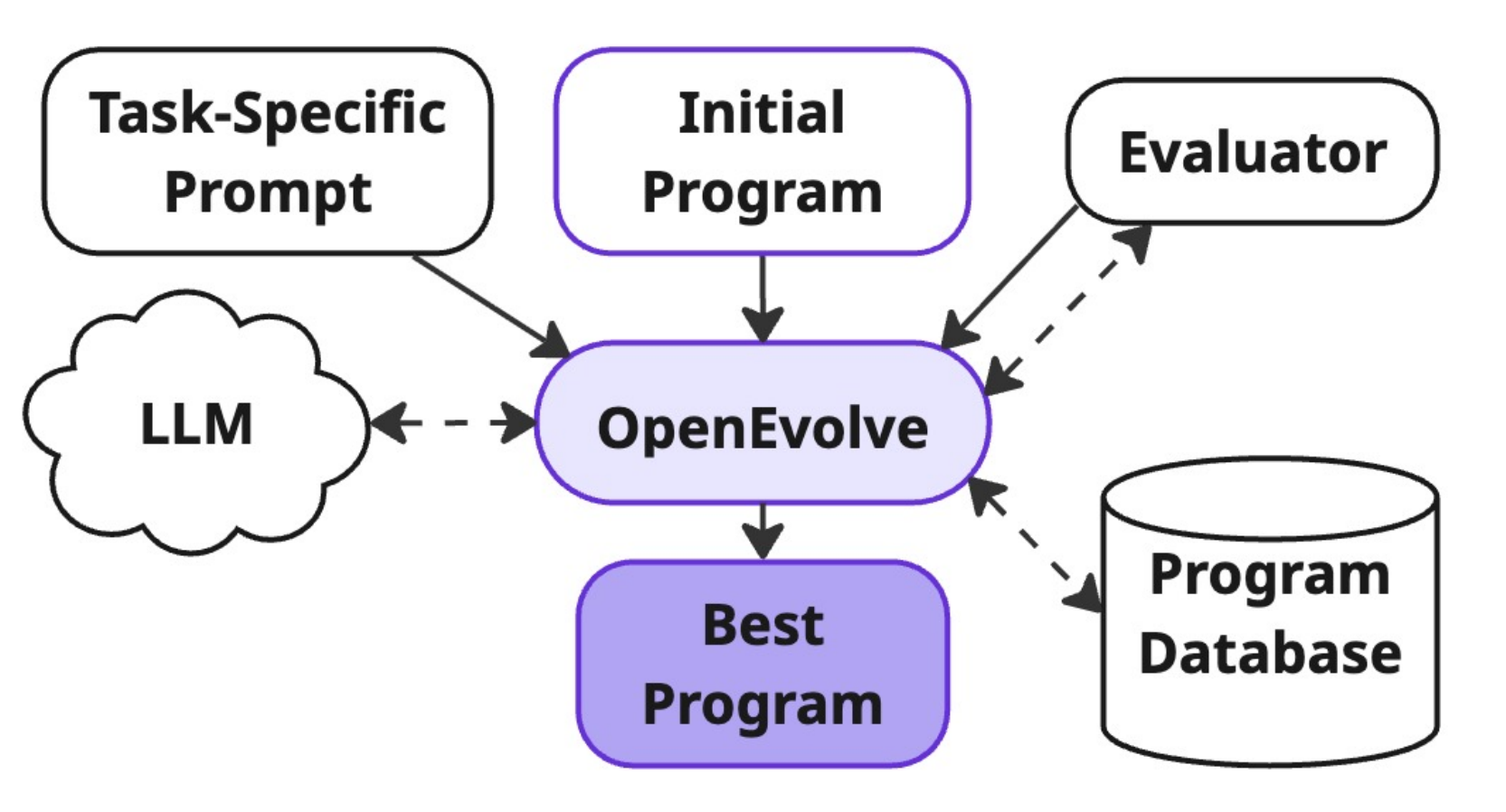}
        \caption{OpenEvolve.}
        \label{fig:openevolve-workflow}
    \end{subfigure}%
    \hfill
    \begin{subfigure}[t]{0.33\textwidth}
        \centering
        \includegraphics[width=\linewidth]{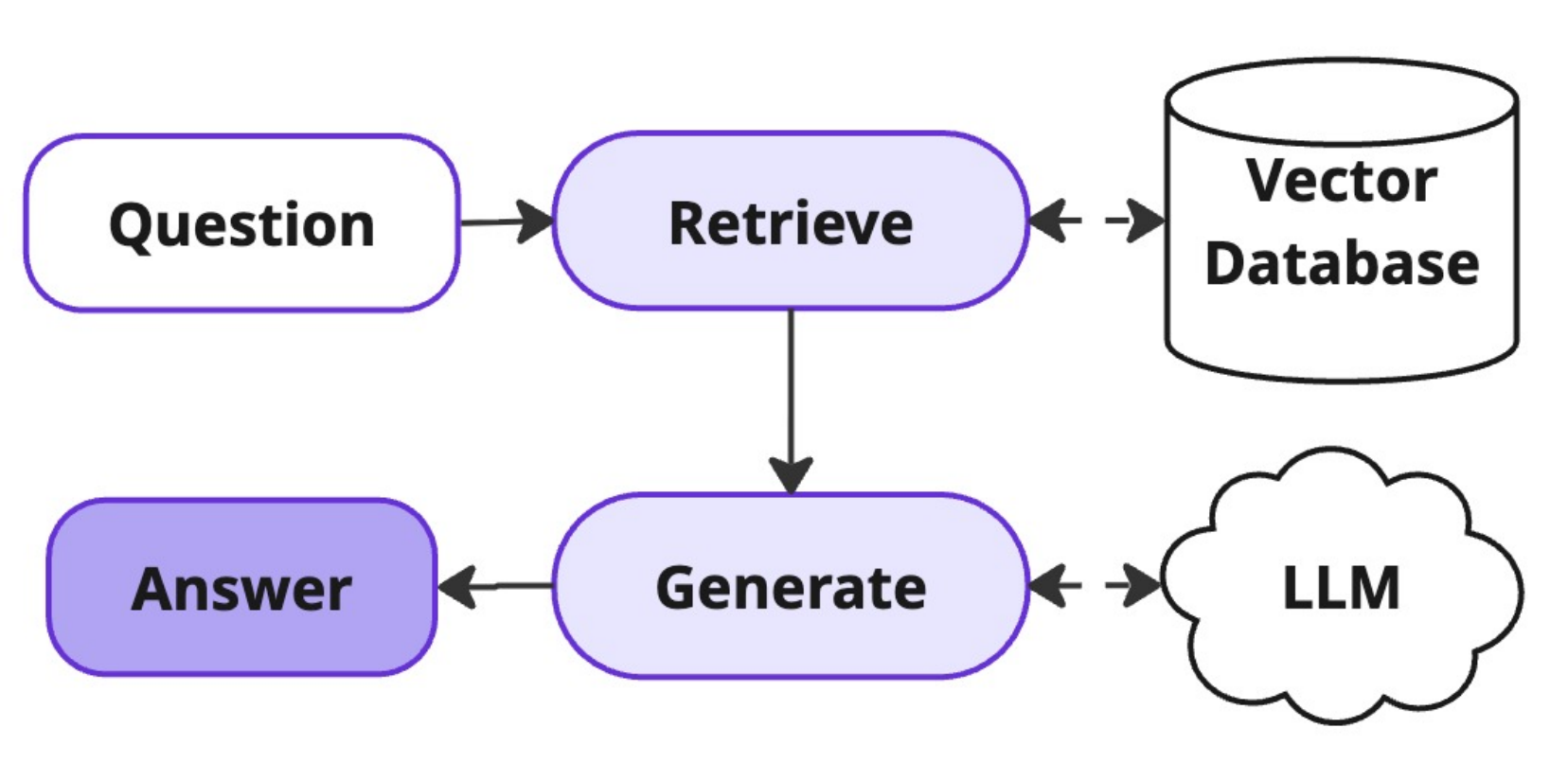}
        \caption{Retrieval-Augmented Generation.}
        \label{fig:rag-workflow}
    \end{subfigure}
    \caption{Overview of three workflows from our benchmark suite.}
    \label{fig:workflows}
\end{figure*}


We study three Compound AI applications: (a) Video Question/Answering (QA), (b) OpenEvolve, and (c) Retrieval-Augmented Generation (RAG). We include these applications because they represent complex, multi-modal industrial workflows~\cite{viva2022romero,zhang2024omagent,novikov2025alphaevolvecodingagentscientific} that impose varied operational demands. Our suite spans multiple data modalities—including text, audio, and video—and covers diverse tasks such as vector retrieval, multi-modal reasoning, and iterative code generation. Furthermore, these applications represent different service-level objectives: Video-QA and RAG function as interactive, latency-sensitive pipelines, whereas OpenEvolve operates as a batch job with flexible timing. This diversity allows us to analyze how different latency targets and heterogeneous execution patterns stress system resources in distinct ways.


\subsection{Video Question/Answering (QA)}


The Video-QA application~\cite{zhang2024omagent}, illustrated in \Cref{fig:video-qa_workflow}, generates answers to user questions about video content.
First, a Video Encoder extracts frames and raw audio, which a Whisper~\cite{whisper} Speech-to-Text (STT) model converts into a text transcript. 
A Gemma-3-27B~\cite{gemmateam2025gemma3technicalreport} multi-modal LLM then generates a response using the audio transcript and video frames.
This application is a representative benchmark with practical utility in manufacturing, security, and forensics~\cite{tang2025videounderstandinglargelanguage,agrawal2024agentic,zhang2024omagent}.


\subsection{OpenEvolve}


OpenEvolve~\cite{openevolve}, illustrated in \Cref{fig:openevolve-workflow}, is an evolutionary agent for algorithm discovery and optimization, mirroring AlphaEvolve~\cite{novikov2025alphaevolvecodingagentscientific} which is used for cluster scheduling and kernel-tuning~\cite{wei2025astramultiagentgpukernel,2025kernelbench,georgiev2025mathematicalexplorationdiscoveryscale}.
It employs a multi-turn feedback loop between a CPU control process and a GPU-based LLM.
The system initializes a program, prompt, and evaluator on the CPU, then constructs an iterative template containing historical performance data.
The LLM generates program variants that are evaluated on the CPU or GPU, with results stored in a database to guide subsequent iterations toward the optimal program.

\subsection{Retrieval-Augmented Generation (RAG)}
Retrieval-Augmented Generation (RAG)~\cite{lewis2020rag} enhances the capabilities of LLMs by integrating relevant information from external sources into the response generation process. Instead of depending solely on a model's pre-trained knowledge, RAG retrieves supporting documents to produce more accurate, reliable, and current outputs. 

\autoref{fig:rag-workflow} that our RAG workflow consists of two primary stages, the Retrieve stage and the Generate stage. In the Retrieve stage, the system converts the user's query to a vector using an embedding model. It then queries a vector database, which stores representations of document chunks, to find the top $k$ most relevant vectors based on similarity. These retrieved documents and the user query are used to prompt the LLM in the Generate phased to produce the final answer. 

For evaluation, we use a Gemma-3-27B model as the generation component, with vLLM handling model execution and Milvus Lite serving as the vector database. For the accuracy–latency tradeoff experiment, documents were embedded using a Text-Embedding-3-Small and split into 2,000-token chunks with a 200-token overlap. The system retrieved the top $k$ chunks, with $k$ ranging from 5 to 30, and was evaluated on the Google FRAMES benchmark across three database configurations: a baseline, a 100-question, and the full 824-question database. For the resource analysis experiment, EmbeddingGemma was employed with 1,000-token chunks and a 100-token overlap against a 3.6 GB vector database, including all 824 benchmark questions.

\subsection{Benchmarking Infrastructure}

We design our infrastructure to be workflow-agnostic, allowing developers to easily integrate new Compound AI applications for cross-stack analysis without extensive refactoring. The infrastructure consists of several modular components focused on deployment, monitoring, and hardware manipulation.

We serve the models using vLLM, which supports a wide range of open-source architectures and provides native instrumentation metrics. For components that require CPU-side flexibility, such as the Whisper speech-to-text model in Video-QA, we use Docker to run prepackaged images. To simulate realistic usage patterns, the suite includes a load generator that issues queries at determined rates with inter-arrival times following the Poisson distribution. 

For monitoring the applications, we use a vLLM monitor, which periodically scrapes the metrics exposed by vLLM. Specifically, to measure CPU metrics, we use Linux's System Activity Reporter (SAR), while for GPU metrics, we use Nvidia's Data Center GPU Manager Interface (DCGMI).
\section{Hardware Implications}

The vast array of available accelerators, including NVIDIA and AMD GPUs, TPUs, offers significant opportunities for optimizing resource allocation in Compound AI systems.
Individual application components respond differently to hardware, and their combination complicates end-to-end configuration decisions.
To study this space, we first analyze how different workloads distribute their execution time between CPU and GPU resources to identify primary system bottlenecks. Next, we evaluate applications across heterogeneous accelerators with varying model parallelism to measure performance and resource use.
Finally, we analyze how hardware-level configurations for specific accelerators impact individual components and end-to-end metrics.

\subsection{Resource Dominance Analysis}
\begin{figure}[h]
\centering
\begin{tikzpicture}
\begin{axis}[
    xbar stacked,
    bar width=18pt,            
    width=0.9\columnwidth,
    height=4cm,
    clip=false,                
    enlarge y limits={abs=0.8cm}, 
    hide x axis,               
    axis y line=left,          
    y axis line style={-},     
    symbolic y coords={OpenEvolve, Video-QA, RAG},
    ytick=data,
    yticklabel style={font=\small},
    xmin=0, xmax=100,
    legend style={
        at={(0.5,-0.2)}, 
        anchor=north, 
        legend columns=-1, 
        draw=none,
        font=\small
    },
    nodes near coords={\pgfmathprintnumber{\pgfplotspointmeta}\%}, 
    every node near coord/.append style={font=\footnotesize\bfseries}
]



\addplot [fill=blue!80!black, draw=white, nodes near coords style={white}] coordinates {
    (8,RAG)
    (62,Video-QA) 
    (82,OpenEvolve) 
};

\addplot [fill=gray!20, draw=gray!40, nodes near coords style={black}] coordinates {
    (92,RAG) 
    (38,Video-QA) 
    (18,OpenEvolve) 
};

\legend{GPU Dominance, CPU Dominance}

\end{axis}
\end{tikzpicture}
\caption{Temporal resource dominance in the end-to-end execution of a single request for each application. Percentages indicate the proportion of total timestamps where the specified resource maintains higher utilization.}
\label{fig:cpu_gpu_breakdown}
\end{figure}
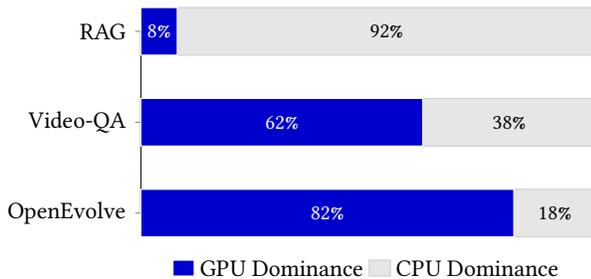

We run all three workloads in our benchmark with the Gemma-3-27B model and periodically measure the CPU and GPU utilization. \autoref{fig:cpu_gpu_breakdown} shows the breakdown of time period where CPU or GPU utilization dominates. While the systems community has historically focused on optimizing GPU execution for AI workloads, our analysis reveals that CPU execution remains a critical component of the end-to-end pipeline. In Retrieval-Augmented Generation (RAG) workflows, the CPU maintains higher utilization for 92\% of the execution time, largely due to the overhead of vector database management and prompt orchestration. Conversely, Video-QA and OpenEvolve exhibit higher GPU dominance at 62\% and 82\%, respectively, reflecting their heavy reliance on multi-modal LLM inference and evolutionary program generation.

\begin{figure}[ht]
  \centering
    \includegraphics[width=1\linewidth]{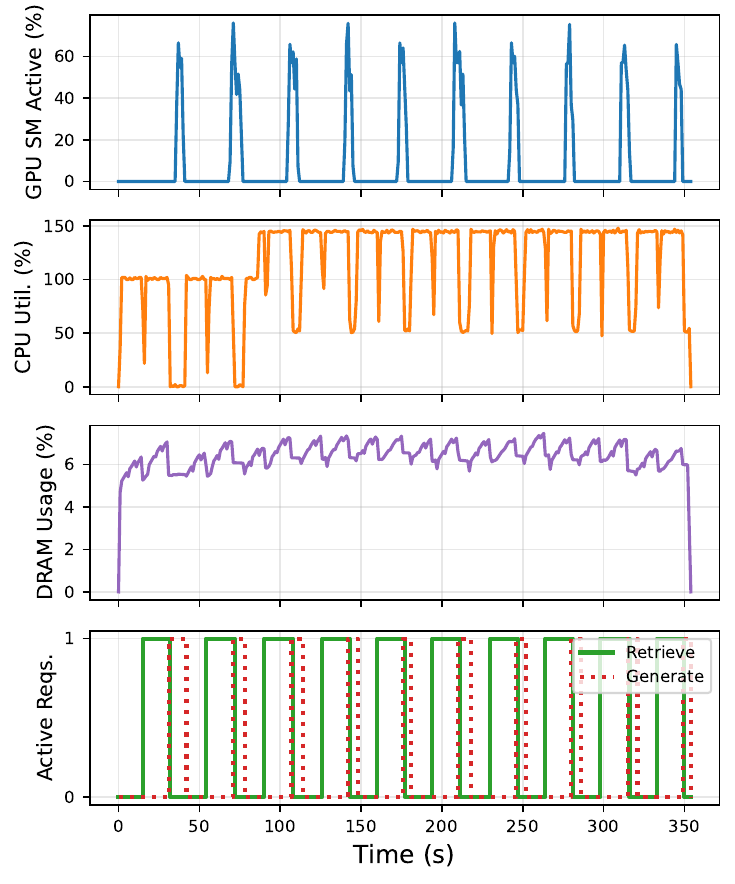}
    \caption{System resource utilization timeline of RAG pipeline execution with sequential requests, when using the Gemma model.}
    \label{fig:rag-workflow-1}
    \vspace{-0.5cm}
\end{figure}

To further understand the temporal patterns of resource dominance, we analyze the system resource utilization timelines during RAG pipeline execution. \autoref{fig:rag-workflow-1} illustrates the resource utilization of RAG pipeline execution when requests are sent sequentially. We observe that GPUs' Streaming Multiprocessors (SM) are activated in a spiky pattern and are idle for the majority of the time, stalled by the retrieve phase, which executes on the CPU.
We also analyze the host DRAM usage over time---it can be observed that the retrieval phase in each request loads up the relevant data into memory for vector search, causing a spike in memory usage.
With a higher load of concurrent retrieval requests and larger databases, this can bottleneck the end-to-end execution, especially if the data does not fit into DRAM and needs to be fetched from a lower tier like remote storage or disk.

\begin{figure}[ht]
  \centering
    \includegraphics[width=1\linewidth]{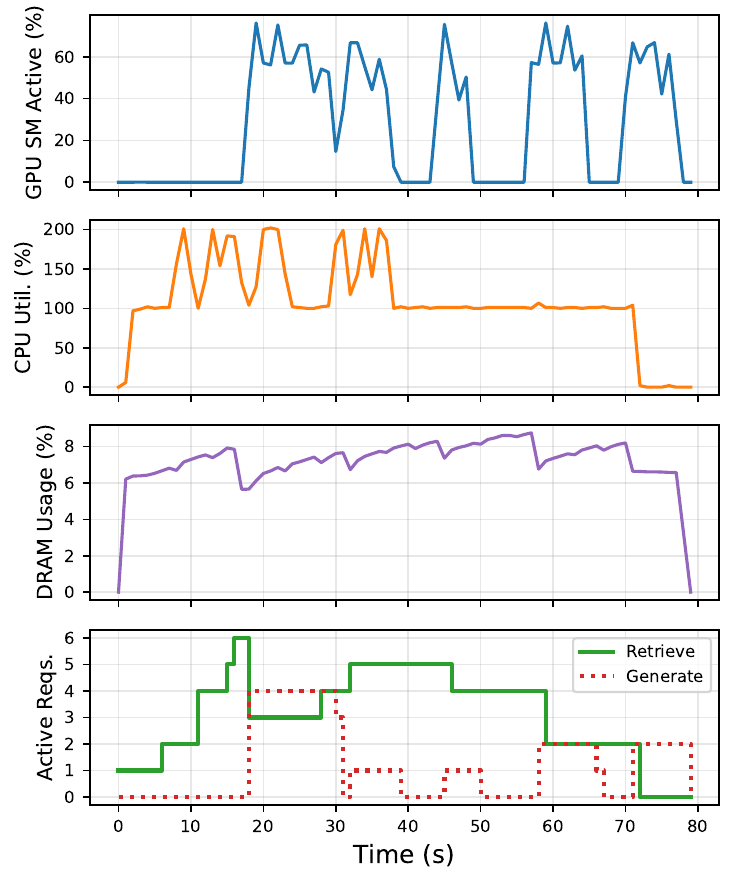}
    \caption{System resource utilization timeline of RAG pipeline execution with Poisson request arrival rate of 0.3, when using the Gemma model..}
    \label{fig:rag-workflow-2}
\end{figure}

Under more complex load conditions, such as a RAG pipeline with a Poisson request arrival rate of 0.3 (\autoref{fig:rag-workflow-2}), overlapping requests cause higher sustained CPU utilization as multiple retrieval stages occur concurrently. The timeline demonstrates that as the active request count for retrieval increases, the GPU's active periods become more frequent; however, significant idle periods remain. These results suggest that system efficiency can be improved by a design that properly pipelines CPU and GPU executions to mitigate bottlenecks.

\takeaway{Optimizing CPU execution is just as vital as optimizing GPU execution for Compound AI applications. Because these applications lack a "one-size-fits-all" hardware setup, system efficiency requires a balanced approach that addresses both accelerator performance and the significant CPU-side bottlenecks.}

\subsection{Accelerator Selection}

\begin{table*}[!h]
\centering
\small
\begin{tabular}{l c c c c c c c}
\toprule
\textbf{Accelerator} & \textbf{TP} & \textbf{Energy} & \textbf{E2E Lat.} & \textbf{P99 Power} & 
\textbf{Price\textsuperscript{$\dagger$}} &
\textbf{Total Cost\textsuperscript{$\dagger$}} & \textbf{Note} \\ 
 & & \small(Wh) & \small(s) & \small(W) & 
 \small(\$/hr) &
 \small(\$) & \\
\midrule

\textbf{NVIDIA L40S} & 1 & -- & -- & -- & 0.47 & -- & \\ 
\small 48GB | 362 TFLOPS & 2 & 250 $\pm$ 70 & 2070 $\pm$ 73 & 321.9 $\pm$ 1.1 & 0.93 & 0.53 & \textit{Min. Power} \\ 
\midrule

\textbf{NVIDIA A100 PCIe} & 1 & 168 $\pm$ 20 & 2292 $\pm$ 285 & 507.0 $\pm$ 2.4 & 0.52 & 0.33 & \textit{Min. Cost} \\
\small 80GB | 312 TFLOPS & 2 & 278 $\pm$ 60 & 2373 $\pm$ 478 & 410.8 $\pm$ 90.7 & 1.04 & 0.69 & \\ 
\midrule

\textbf{NVIDIA H100 SXM} & 1 & 144 $\pm$ 20 & 1531 $\pm$ 101 & 657.1 $\pm$ 18.2 & 1.56 & 0.66 & \\ 
\small 80GB | 1979 TFLOPS & 2 & 248 $\pm$ 50 & 1552 $\pm$ 283 & 553.0 $\pm$ 64.7 & 3.12 & 1.35 & \\ 
\midrule

\textbf{NVIDIA H200 SXM} & 1 & 132 $\pm$ 20 & 1511 $\pm$ 215 & 587.2 $\pm$ 67.2 & 2.19 & 0.92 & \textit{Min. Energy} \\ 
\small 141GB | 1979 TFLOPS & 2 & 190 $\pm$ 10 & 1307 $\pm$ 64 & 423.4 $\pm$ 10.9 & 4.38 & 1.59 & \textit{Min. Latency} \\
\bottomrule
\addlinespace[0.5ex]
\multicolumn{6}{l}{\footnotesize $^\dagger$ Based on the on-demand spot instance pricing offered by Vast.ai~\cite{VastPricing2026}.}
\end{tabular}
\caption{OpenEvolve (Circle Packing) on different accelerators. Metrics reported as mean $\pm$ std. deviation of three runs.}
\label{tab:gpu_prompt_default_comparison}
\end{table*}

The choice of accelerators depends of the trade-off among latency, power, energy and cost acceptable to the application user and operator.
To study this trade-off space, we use OpenEvolve for the Circle Packing task with a Qwen3-Coder-30B~\cite{qwen_qwen3_coder_next_tech_report} model across various NVIDIA GPUs~\cite{nvidia_a100,nvidia_h100,nvidia_v100}.
\Cref{tab:gpu_prompt_default_comparison} reveals that the NVIDIA H200 with Tensor Parallelism (TP) 2 achieves the lowest latency.
However, the H200 with TP 1 proves more energy-efficient; while it takes approximately 15.6\% longer to execute than the TP 2 setup, it consumes roughly 30.5\% less energy by avoiding the power overhead of a second GPU.

Power budgets also dictate hardware selection, as datacenter racks often impose strict limits.
In these experiments, the NVIDIA L40S demonstrates the lowest $99^{th}$ percentile power consumption but increases latency by approximately 58.4\% compared to the fastest setup. From a cost perspective, the NVIDIA A100 with TP 1 is the most economical. Although the L40S has a lower hourly price, its limited memory requires two instances to run the model, making it approximately 60.6\% more expensive than a single A100.

\takeaway{There is no \emph{single} optimal configuration for the workflow; each choice requires balancing trade-offs. Power-efficient GPUs may consume more total energy due to longer runtimes, while scaling up GPU counts can result in diminishing returns, increasing cost with limited performance gains. The serving platform must consider these multi-dimensional characteristics and the SLOs to \emph{right-size} the resources.}

\subsection{Configuration Sensitivity}

\begin{figure*}[!h]
    \centering
    \includegraphics[width=\textwidth]{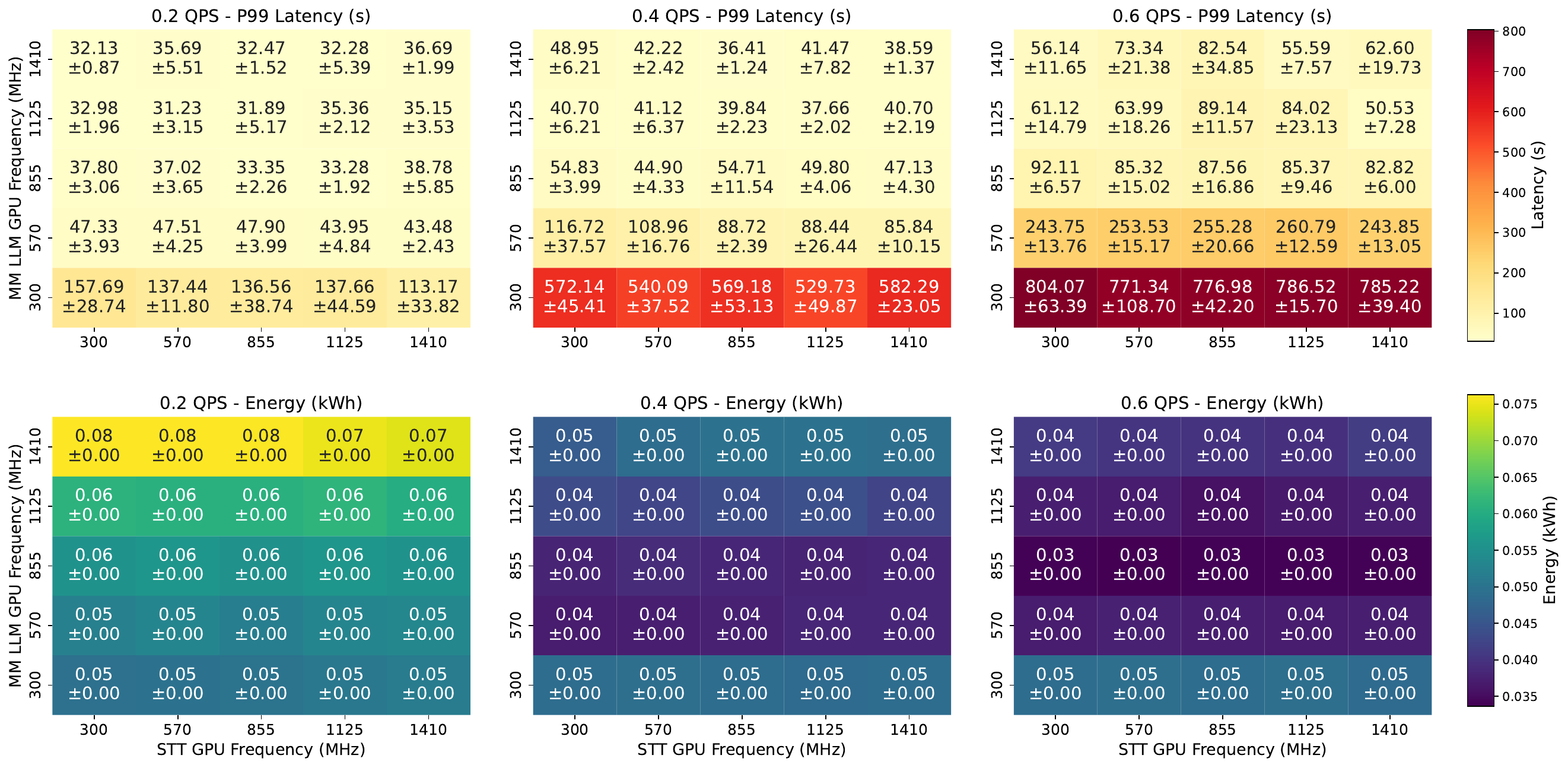}
    \caption{Video-QA sensitivity to per-component GPU frequencies and impact on energy and tail latency under varying load.}
    \label{fig:lat+energy_combined}
\end{figure*}

We study the sensitivity of individual components of a Compound AI application to changes in hardware configurations.
We use the Video-QA application, where we allocate one NVIDIA A100 GPU each for the multi-modal LLM (MM LLM) model and a speech-to-text (STT) model.
We change the Streaming Multiprocessor (SM) frequency of the GPUs using Nvidia's System Management Interface (nvidia-smi) and observe its impact on end-to-end application metrics.

\subsubsection{Latency vs. Energy Consumption}
\Cref{fig:lat+energy_combined} shows the heatmaps of energy consumption and 99-th percentile latency for different configurations of SM frequencies for the MM LLM and STT model.

At low loads -- 0.1 queries per second (QPS) with Poisson arrivals -- the STT model does not bottleneck the system, so increasing its frequency yields no significant latency gains.
Conversely, the computationally expensive MM LLM shows high sensitivity; increasing its frequency from 300~MHz to 1125~MHz halves tail latency.
Capping the LLM at 1125~MHz and the STT at 300~MHz reduces energy consumption by roughly 30\% compared to maximum frequency settings.

At high loads (0.4~QPS), a low frequency for MM LLM causes tail latency to spike to 16$\times$ the minimum possible value.
At medium load (0.2~QPS), operating the STT's GPU at the highest frequency only makes sense if the multi-modal model GPU is running at lower frequencies like 300~MHz and 570~MHz.
For example, when the MM LLM runs at 300~MHz, increasing the STT GPU’s frequency from 300~MHz to 1410~MHz results in a 28.2\% reduction in tail latency, while maintaining approximately the same energy consumption.
At higher GPU frequencies for the MM LLM, the impact of a higher STT GPU frequency becomes unnoticeable.
These results confirm that not all components benefit from maximum frequency, and dynamically adjusting these settings based on load and sensitivity improves resource efficiency without violating latency targets.

\takeaway{Individual components in an applications demonstrate varied sensitivity to hardare configuration changes.
For highest resource-efficiency while also satisfying SLOs, the serving platform must \emph{dynamically} adjust hardware-level knobs (\eg{} GPU frequency) of \emph{each} component after considering its impact on the \emph{end-to-end} application behavior.}


\subsubsection{Power Consumption}

\begin{figure}[t]
    \centering
    \includegraphics[width=\linewidth]{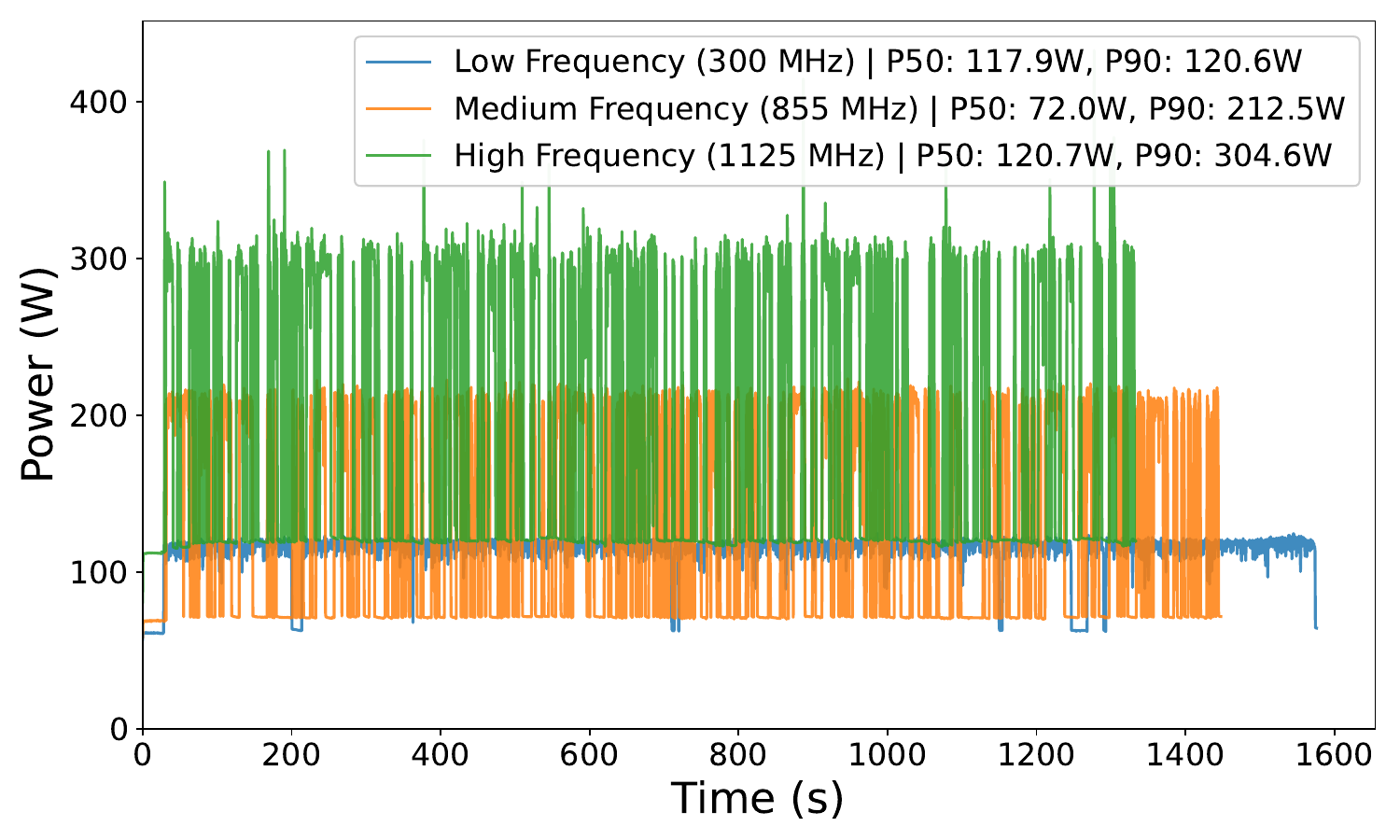}
    \caption{Video-QA MM LLM GPU's power draw at different frequency configurations.}
    \label{fig:comparison_p50_p90_freq}
\end{figure}

AI applications greatly stress the power infrastructure at various levels, from the rack to the power grid~\cite{stojkovic2025tapasthermalpowerawarescheduling,choukse2025powerstabilizationaitraining}.
It is important to appropriately provision power and limit the power draw without compromising user-level performance.
For Compound AI applications, this means understanding carefully how power should be apportioned between different components.
We study this behavior using the Video-QA application by keeping a consistent load (0.2~QPS) and varying only the frequency of the multi-modal model's GPU without altering the frequency of the STT GPU.
\Cref{fig:comparison_p50_p90_freq} depicts the power draw over time and reveals an interesting trend: average power consumption increases as frequency rises from 300 to 855 MHz, but then drops at 1125 MHz.
At low frequency (300~MHz), the workload draws a moderately low power consistently (50-th percentile 117~W and 90-th percentile 120~W) without much variation.
At medium frequency (855~MHz), the power draw demonstrates more fluctuation.
It often goes down close to the GPU idle-power with an average power draw of $\approx$~72~W but frequently touching up to about 212~W.
At high frequency (1125~MHz), the workload is able to draw much higher power, $\approx$~304~W, for short bursts of time while maintaining a higher average of $\approx$~120 W to finish the workload faster.
As the frequency is increased from low to high, the end-to-end execution time falls from $\approx$~1600 to $\approx$~1300~s.

This shows that the trend in average power vs peak power usage can vary greatly for individual components in an application based on the operating frequency of the accelerators and interplay with other components.
Deciding how to configure the individual accelerators may depend on various factors.
For example, if a provider can tolerate higher peak power draw in short bursts, then a higher frequency may be acceptable.
If, on the other hand, a moderate power draw can be sustained but large swings in power cannot be tolerated (\eg{} due to power grid limitations~\cite{choukse2025powerstabilizationaitraining}) then a medium frequency may be the way to go.

\takeaway{Individual components in an applications may exhibit drastically different power profiles depending on hardware configuration.
Datacenter power management software must consider, both, power constraints and the sensitivity of each component to different configurations in order to apportion power budget and meet performance objectives.}
\section{Software Implications}

While hardware selection and configuration provide the foundation for system performance, the high-level orchestration and internal logic of the software pipeline are equally critical to the efficiency of Compound AI applications. Notably, insights derived from software analysis can often be translated into practice more rapidly than hardware-level changes, as they do not depend on the slow fabrication cycles typical of physical chips. By analyzing the software stack, we can reveal opportunities that leverage application semantics such as prompt structure, data routing, and retrieval parameters to improve end-to-end application and system metrics.

\subsection{Application Configuration Tuning}

\begin{figure}[h]
  \centering
    \includegraphics[width=1\linewidth]{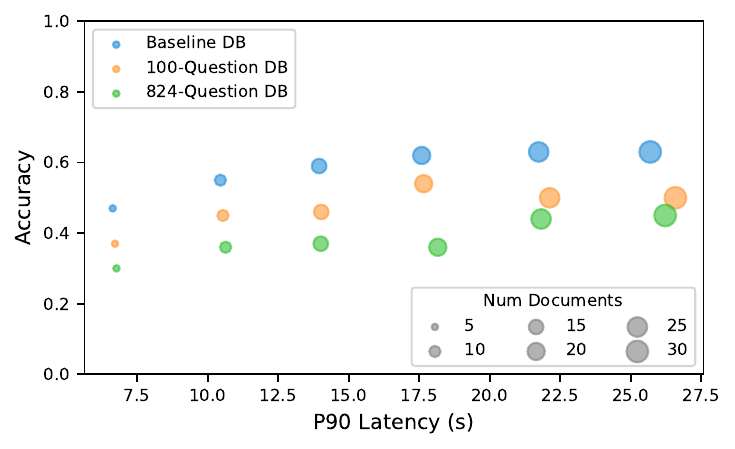}
    \caption{Accuracy versus 90th percentile latency trade-off in RAG application for varying number of retrieved documents.}
    \label{fig:rag-workflow-tradeoff}
\end{figure}

The performance of Compound AI applications is not solely dependent on hardware; it is also highly sensitive to the internal configurations of the software pipeline. In this section, we evaluate the trade-off between accuracy and tail latency within the Retrieval-Augmented Generation (RAG) workflow. We specifically examine the impact of the retrieval parameter $k$, the number of document chunks returned by the Milvus vector database, using the Google FRAMES benchmark and OpenAI's text-embedding-3-small model.

\autoref{fig:rag-workflow-tradeoff} showcases the accuracy and tail latency observed for the RAG workflow when varying the number of documents ($k$) returned by the retrieval stage. When $k$ is 5, the workflow has low $90^{th}$ percentile latency but with low accuracy. As we increase the value of $k$ from 5 to 20, both accuracy and tail latency increase. However, on increasing $k$ further to 30, the tail latency increases with no increase in accuracy. Hence, increasing number of retrieved documents beyond 20 is not useful. 

\takeaway{Accuracy must be treated as a first-class metric in hardware-software co-design. System designers should balance user-level accuracy with strict latency targets.}

\subsection{Cache Management Strategies}

\begin{table*}[!h]
\centering
\small 
\begin{tabular}{lcccccccc}
\toprule
 & & \multicolumn{4}{c}{\textbf{Raw Measurements}} & \multicolumn{2}{c}{\textbf{Normalized Comparison}} \\
\cmidrule(lr){3-6} \cmidrule(lr){7-8}
\textbf{Model} & \textbf{Opt.} & \textbf{E2E Lat.} & \textbf{Energy} & \textbf{Score} & \textbf{KV Hit} & \textbf{Energy to reach} & \textbf{Score at} \\ 
 & & (s) & (kWh) & & (\%) & \textbf{Min Score$^\dagger$} & \textbf{Min Latency$^\ddagger$} \\
\midrule
\multirow{2}{*}{\texttt{Gemma-3-27B-Instruct}} 
    & No  & 6870 & 0.668 & 0.6068 & 1.58 & 0.604 & 0.6068 \\
    & Yes & 6275 & 0.588 & 0.9435 & 17.75 & 0.191 & 0.9435 \\ 
\midrule
\multirow{2}{*}{\texttt{Qwen3-Coder-30B-A3B-Instruct}} 
    & No  & 3486 & 0.304 & 0.8499 & 4.54 & 0.248 & 0.8499 \\ 
    & Yes & 3365 & 0.301 & 0.9378 & 28.88 & 0.075 & 0.9378 \\ 
\bottomrule
\addlinespace[0.5ex]
\multicolumn{8}{l}{\footnotesize $^\dagger$ \textbf{Min Score}: The lowest final score between Opt/Unopt (Gemma: 0.6068, Qwen3: 0.8499).} \\
\multicolumn{8}{l}{\footnotesize $^\ddagger$ \textbf{Min Latency}: The lowest final E2E latency between Opt/Unopt (Gemma: 6275s, Qwen3: 3365s).}
\end{tabular}
\caption{Performance comparisons of prompt optimization on NVIDIA A100.}
\label{tab:gpu_prompt_opt_benefit_comparisons}
\end{table*}

LLMs being autoregressive lend themselves to the caching of Key-Value (KV) pairs, a fundamental part of the attention mechanism~\cite{pope2022efficientlyscalingtransformerinference}.
This KV cache management has emerged as a critical optimization technique for accelerating LLM inference by reducing redundant computations and improving memory utilization~\cite{kwon2023efficientmemorymanagementlarge}.
Moreover, multi-modal LLMs may also benefit from multi-modal (MM) caching to prevent the repeated processing of multimedia data (images, audio, etc.) across different requests.Balancing limited cache capacity across these different types of caches is crucial for effective use of High Bandwidth Memory (HBM) on accelerators.

Our study reveals that Compound AI applications suffer from low cache hit rates. First, we discuss our experiments that demonstrate the low cache hit rates of OpenEvolve and Video-QA applications. Next, we propose system-level optimizations that can dramatically improve cache hit rates. 

\begin{figure*}[!h]
  \centering
  \begin{subfigure}[b]{0.49\linewidth}
    \includegraphics[width=\textwidth]{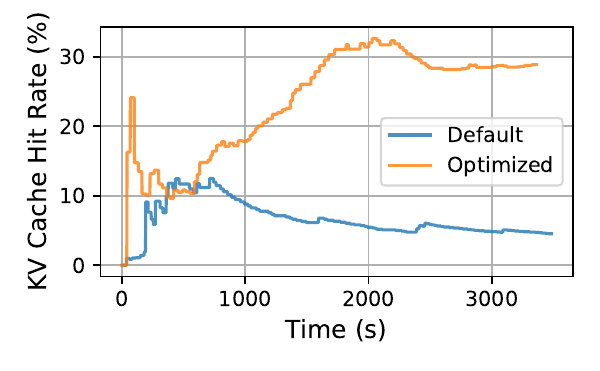}
    \caption{KV cache hit rate.}
    \label{fig:openevolve-rust-sort-job-hit-rate}
  \end{subfigure}\hfill
  \begin{subfigure}[b]{0.49\linewidth}
    \includegraphics[width=\textwidth]{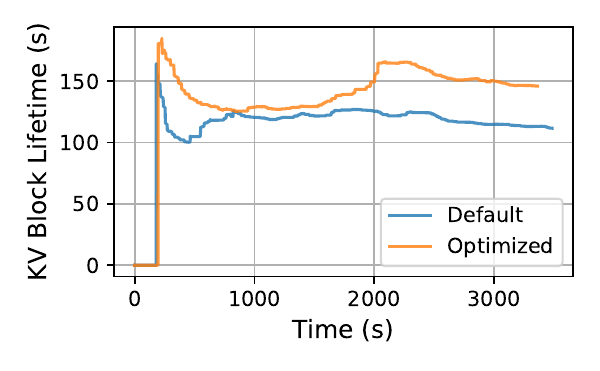}
    \caption{KV cache average block lifetime.}
    \label{fig:openevolve-rust-sort-kv-block-lifetime}
  \end{subfigure}
  \caption{Comparison of KV cache usage for OpenEvolve between the default implementation vs. prompt-reordering (optimized).}
  \label{fig:openevolve-rust-sort}
\end{figure*}

We run OpenEvolve to optimize an adaptive sorting algorithm using the Qwen3-Coder-30B~\cite{qwen_qwen3_coder_next_tech_report} and Gemma3-27B~\cite{gemmateam2025gemma3technicalreport} models for 100 iterations.
\autoref{fig:openevolve-rust-sort-job-hit-rate} shows that the standard prompt template (Default line in the plot) inadvertently sabotages KV cache reuse. By placing frequently changing sampled data at the start of a prompt, the system prevents the serving engine from utilizing prefix caching, as even a single token change at the beginning of a string invalidates the subsequent cache for that request and subsequent requests with potentially the same prefix.

\begin{figure}[!h]
    \centering
    \includegraphics[width=1\linewidth]{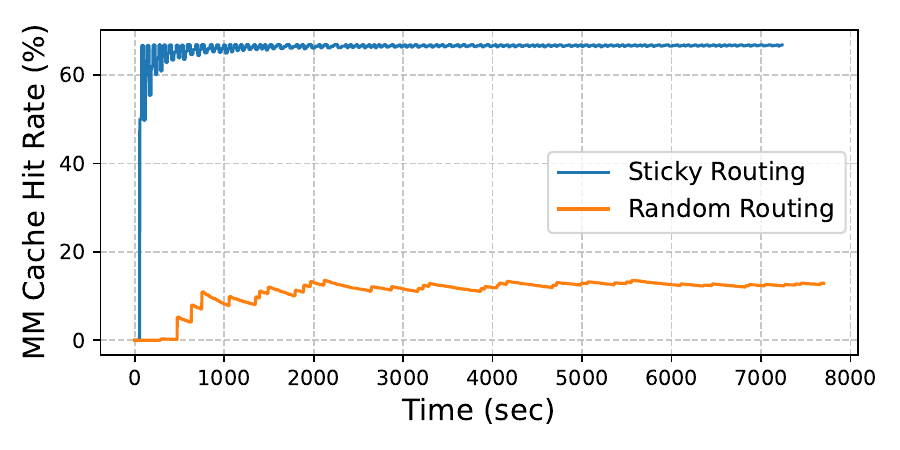}
    \caption{MM cache usage for Video-QA for random routing and sticky routing.}
    \label{fig:vid_mm_plot}
\end{figure}

Next, we analyze the cache behavior of the Video-QA workload, a multi-modal application. In this context, the brevity of the questions ensures that the MM cache efficiency -- rather than text processing -- is the primary driver of performance. We used two A100 GPUs with a 10GB MM cache to process videos of 100 frames each, with STT enabled. We maintained a consistent 50\% GPU utilization to prevent compute saturation. Conventionally, requests are randomly routed to either GPU, which leads to recomputation of contents in the MM cache if the video content of the request is not present in the cache. \autoref{fig:vid_mm_plot} demonstrates that such a random routing strategy leads to an abysmally low MM cache hit rate of 13\%.%

\takeaway{Compound AI applications exhibit poor KV and MM cache hit rates.}

Now that we have demonstrated the poor cache hit rate of Compound AI applications, we propose optimizations that will enable systems to manage caches better.

\subsubsection{Cache-aware Prompt Optimization}

\definecolor{redframe}{RGB}{180, 40, 30}
\definecolor{greenframe}{RGB}{60, 140, 60}
\definecolor{redbg}{RGB}{250, 230, 230}
\definecolor{greenbg}{RGB}{230, 250, 230}

\newcommand{\hlred}[1]{{\setlength{\fboxsep}{0.5pt}\colorbox{redbg}{\hspace{1pt}#1\hspace{1pt}}}}
\newcommand{\hlgreen}[1]{{\setlength{\fboxsep}{0.5pt}\colorbox{greenbg}{\hspace{1pt}#1\hspace{1pt}}}}

\begin{figure}[h]
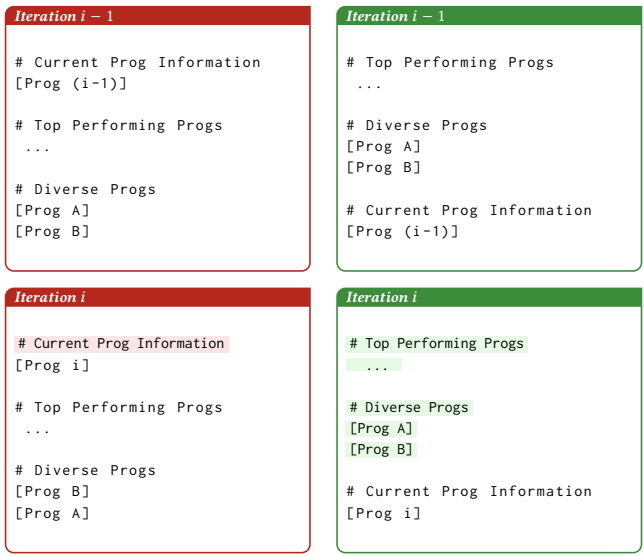

\centering
\begin{subfigure}[t]{0.48\columnwidth}
    \centering
    \begin{tcolorbox}[title={Iteration $i-1$}, colframe=redframe, colback=white, fonttitle=\tiny\bfseries\itshape, sharp corners=northeast, boxrule=0.6pt, boxsep=1pt, left=2pt, right=2pt, after skip=2mm]
    \begin{lstlisting}[basicstyle=\ttfamily\tiny, escapeinside={(*@}{@*)}]
# Current Prog Information
[Prog (i-1)]

# Top Performing Progs
 ...

# Diverse Progs
[Prog A]
[Prog B]
    \end{lstlisting}
    \end{tcolorbox}

    \begin{tcolorbox}[title={Iteration $i$}, colframe=redframe, colback=white, fonttitle=\tiny\bfseries\itshape, sharp corners=northeast, boxrule=0.6pt, boxsep=1pt, left=2pt, right=2pt]
    \begin{lstlisting}[basicstyle=\ttfamily\tiny, escapeinside={(*@}{@*)}]
(*@\hlred{\# Current Prog Information}@*)
[Prog i]

# Top Performing Progs
 ...

# Diverse Progs
[Prog B]
[Prog A]
    \end{lstlisting}
    \end{tcolorbox}
    \caption{Default Prompt}
    \label{fig:prompt_default}
\end{subfigure}
\hfill
\begin{subfigure}[t]{0.48\columnwidth}
    \centering
    \begin{tcolorbox}[title={Iteration $i-1$}, colframe=greenframe, colback=white, fonttitle=\tiny\bfseries\itshape, sharp corners=northeast, boxrule=0.6pt, boxsep=1pt, left=2pt, right=2pt, after skip=2mm]
    \begin{lstlisting}[basicstyle=\ttfamily\tiny, escapeinside={(*@}{@*)}]
# Top Performing Progs
 ...

# Diverse Progs
[Prog A]
[Prog B]

# Current Prog Information
[Prog (i-1)]
    \end{lstlisting}
    \end{tcolorbox}

    \begin{tcolorbox}[title={Iteration $i$}, colframe=greenframe, colback=white, fonttitle=\tiny\bfseries\itshape, sharp corners=northeast, boxrule=0.6pt, boxsep=1pt, left=2pt, right=2pt]
    \begin{lstlisting}[basicstyle=\ttfamily\tiny, escapeinside={(*@}{@*)}]
(*@\hlgreen{\# Top Performing Progs}@*)
(*@\hlgreen{\ \phantom{A}... }@*)

(*@\hlgreen{\# Diverse Progs}@*)
(*@\hlgreen{[Prog A]}@*)
(*@\hlgreen{[Prog B]}@*)

# Current Prog Information
[Prog i]
    \end{lstlisting}
    \end{tcolorbox}
    \caption{Optimized Prompt}
    \label{fig:prompt_optimized}
\end{subfigure}

\caption{ Highlights demonstrate the segments that result in cache hits. Optimized prompt improves cache hits via static-to-dynamic ordering.}
\label{fig:openevolve_prompt_examples}
\end{figure}

OpenEvolve is unable to reap the benefits of prefix caching, as frequently changing data is placed at the start of the prompt.  To resolve this, we redesigned the prompt generation logic to prioritize a \emph{static-to-dynamic sequencing} principle, as illustrated in \Cref{fig:openevolve_prompt_examples}.
By moving slowly-changing information, such as top-performing programs, to the beginning of the template and implementing deterministic sorting for dynamic elements (based on database insertion order), we ensured that identical sets of programs consistently produce identical prefixes. This modification is universal; it generalizes to any multi-turn LLM task where maintaining a stable prefix can significantly reduce redundant computation.

The impact of this restructuring is visible in both model performance and hardware efficiency. As illustrated in the \Cref{tab:gpu_prompt_opt_benefit_comparisons}, the optimized prompt improved KV cache hit rate by 16\% and 24\% for Gemma and Qwen models, respectively.
This improved the end-to-end latency up to 8\% and the energy usage up to 12\%.
Moreover, it allowed the Qwen model to surpass a 0.80 performance threshold by the tenth iteration, eventually peaking at 0.94.
The default prompt, conversely, remained stagnant for 80 iterations due to the lack of consistent context.

The performance gap is directly tied to the prefix hit rate.
\Cref{fig:openevolve-rust-sort-job-hit-rate} shows that the default prompt's hit rate decays as the program database grows more diverse, while the optimized prompt stabilizes.
The plot shows a long-term hit rate plateau of approximately $28\%$, which is expected based on a request structure where the top 4 of 14 programs remain consistent at the start of the sequence.
Furthermore, \Cref{fig:openevolve-rust-sort-kv-block-lifetime} shows that this optimization significantly extends KV block lifetime.
In a steady state, the optimized prompt maintains a higher average block lifetime because a notable portion of the cache is dedicated to a persistent prefix.
While non-prefix blocks are evicted regularly, the retention of these anchor blocks reduces the frequency of total cache misses.

\takeaway{Prompt optimization frameworks (\eg{} DSPy~\cite{dspy_paper}) should optimize not only for program accuracy, but also for resource-efficiency.
This can greatly maximize cache hit rates and hardware throughput without altering the underlying program semantics.}

\subsubsection{Cache-Aware Routing}

Video-QA suffered from poor MM cache hit rate because the requests were routed to GPUs randomly without considering the state of their caches. To resolve this we propose a sticky routing strategy. Sticky routing directs all requests associated with a specific video to the same machine. We simulated this by processing requests for the same video sequentially, thereby maximizing MM cache reuse. Conversely, random routing distributes requests across the cluster without regard for the underlying video data. In this scenario, the MM cache size becomes a critical bottleneck, as video frames are often evicted to make room for interleaved requests before the next query for that same video arrives.

\Cref{fig:vid_mm_plot} reveals a stark contrast in efficiency: sticky routing achieved a 67\% MM cache hit rate -- a logical result given that our dataset contains three requests per video, where only the initial request must incur a cold start. However, under random routing, the hit rate dropped to just 13\% due to aggressive frame eviction. These cache misses resulted in significant performance penalties. Random routing increased $25^{th}$ percentile latency to 11.924 seconds, a 23.8\% spike compared to sticky routing, while $50^{th}$ and $95^{th}$ percentile latencies rose by 16.9\% and 7.9\% respectively.

\takeaway{Modern serving infrastructure processing high-dimensional data, such as video frames, should perform cache-aware routing to boost MM cache hits and reduce end-to-end execution time and resource consumption.}

\subsubsection{Object-level Memory Signaling} 

While ``stickiness'' at the routing level is a powerful optimization, it functions as an external orchestration policy. To reach the next level of resource efficiency, the system can leverage explicit memory signaling to provide the serving stack with deeper insight into data longevity.

In the future, allowing applications to specify hints about object reuse can unlock better cache usage, similar to how it is done in traditional systems programming.
For example, Linux exposes the \texttt{madvise}~\cite{madvise2} syscall to provide the kernel with directions on how memory will be used, enabling more intelligent management.
Integrating similar interfaces into AI serving stacks like vLLM~\cite{kwon2023efficient} would allow for granular, object-level control.
By signaling which components are static or frequently reused, the system can prioritize and retain them in memory even when sticky routing is not viable, ensuring high-dimensional data like video frames remain accessible for subsequent queries.
This can also allow applications to signal which type of modalities they will be using more, thus apportioning expensive memory resources (\eg{} HBM in GPUs) to the \emph{right} type of cache (\eg{} more memory allocated to MM cache for Video-QA, compared to the KV cache).

\takeaway{Serving engines should provide granular, object-level control over the caches to allow applications to help better utilize expensive memory resources.}



\section{Conclusion and Future Work}
Our cross-stack benchmark suite helps illuminate the large configuration space across the hardware-software stack.
Through our characterization and analysis, we show the importance of designing the entire stack in-tandem---considering the sensitivity of sub-components in an application to hardware knobs, impact of application semantics on resource efficiency via proper software abstractions and management policies.
These decisions have implications on the cost, accuracy, latency, power and energy consumption of individual application components as well as on end-to-end execution.
In the future, we want to focus on exposing more interfaces across different layers of the stack that can capture application semantics and perform better runtime adaptation.
Moving forward, we hope our work can be the basis of a standardized benchmark for the systems community to perform better hardware-software co-design of the datacenter infrastructure required for serving Compound AI applications.

\balance
\bibliographystyle{ACM-Reference-Format}
\bibliography{paper}

\end{document}